\documentclass[titlepage,pallis,twoside]{wpallis}
\usepackage{fancyh}
\usepackage[l]{floatflt}
\usepackage{epsfig}
\usepackage{amssymb}
\usepackage{latexsym}
\usepackage{times}
\usepackage{slashed,cite}
\usepackage{upgreek}
\numberwithin{equation}{section}

\usepackage{amsmath}
\usepackage{amsfonts}
\usepackage{amsbsy}
\usepackage{amscd}
\usepackage{bbm}

\newcommand{\bal}{\begin{align}}
\newcommand{\eal}{\end{align}}
\newcommand{\beqs}{\begin{subequations}}
\newcommand{\eeqs}{\end{subequations}}
\newcommand{\ec}{\end{center}}
\newcommand{\bec}{\begin{center}}

\newcommand{\eem}{\end{matrix}}
\newcommand{\bem}{\begin{matrix}}
\newcommand{\eeq}{\end{equation}}
\newcommand{\beq}{\begin{equation}}
\newcommand{\ba}{\begin{array}}
\newcommand{\ea}{\end{array}}
\newcommand{\bea}{\begin{eqnarray}}
\newcommand{\eea}{\end{eqnarray}}
\newcommand{\baq}{\begin{eqnarray}}
\newcommand{\eaq}{\end{eqnarray}}

\newcommand\eqs[2]{Eqs.~(\ref{#1}) and (\ref{#2})}

\newcommand\eqss[3]{Eqs.~(\ref{#1}), (\ref{#2}) and (\ref{#3})}

\newcommand{\ftn}{\footnotesize}

\newcommand{\GeV}{{\mbox{\rm GeV}}}

\newcommand{\sFref}[2]{Fig.~\ref{#1}-{\ftn\sf ({#2})}}
\newcommand{\sEref}[2]{Eq.~(\ref{#1}{\ftn\sf {#2}})}

\newcommand{\srEref}[2]{Eq.~({\small\sf R.1}-{#1}{\ftn\sf {#2}})}
\newcommand{\rEref}[1]{Eq.~({\small\sf R.1}-{#1})}

\newcommand{\rFref}[1]{Fig.~{\small\sf R.1}-{#1}}
\newcommand{\rTref}[1]{Table~{\small\sf R.1}-{#1}}
\newcommand\reqs[2]{Eqs.~({\small\sf R.1}-{#1}) and ({\small\sf R.1}-{#2})}

\newcommand{\rSref}[1]{Sec.~{\small\sf R.1}-{#1}}

\newcommand{\etal}{{\it et al.\/}}

\def\to{\rightarrow}

\def\lf{\left(}
\def\rg{\right)}
\newcommand\vev[1]{\langle {#1} \rangle}

\newcommand{\Vhi}{\ensuremath{\widehat V_{\rm IG}}}
\newcommand{\Hhi}{\ensuremath{\widehat H_{\rm IG}}}
\newcommand{\Ohi}{\ensuremath{\Omega}}
\newcommand{\Omg}{\ensuremath{\Omega}}
\newcommand{\Khi}{\ensuremath{K}}
\newcommand{\Whi}{\ensuremath{W}}
\newcommand{\Vhio}{\ensuremath{\widehat V_{\rm IG0}}}

\newcommand{\mP}{\ensuremath{m_{\rm P}}}

\newcommand{\Qef}{\ensuremath{\Lambda_{\rm UV}}}

\def\openone{\leavevmode\hbox{\small1\kern-3.8pt\normalsize1}}

\newcommand{\ft}{\ensuremath{f_{n}}}
\newcommand{\kx}{\ensuremath{k_S}}

\newcommand{\Fcr}{\ensuremath{\Omega_{\rm H}}}
\newcommand{\fr}{\ensuremath{f_{\cal R}}}
\newcommand{\fk}{\ensuremath{\Omega_{\rm H}}}

\newcommand{\re}{\ensuremath{e_m}}

\newcommand{\fpp}{\ensuremath{f_{2}}}
\newcommand{\fsp}{\ensuremath{f_{S\Phi}}}
\newcommand{\Fk}{\ensuremath{\Omega_{\rm K}}}
\newcommand{\ks}{\ensuremath{k_S}}
\newcommand{\ksp}{\ensuremath{k_{S\Phi}}}
\newcommand{\kpp}{\ensuremath{k_{\Phi}}}

\newcommand{\ck}{\ensuremath{c_{\cal R}}}

\newcommand{\msn}{\ensuremath{\what m_{\rm \dph}}}

\newcommand{\ns}{\ensuremath{n_{\rm s}}}
\newcommand{\as}{\ensuremath{a_{\rm s}}}
\newcommand{\As}{\ensuremath{A_{\rm s}}}
\newcommand{\rcc}{\ensuremath{\mathcal{R}}}
\newcommand{\rce}{\ensuremath{\widehat{\mathcal{R}}}}
\newcommand{\Ve}{\ensuremath{\widehat{V}}}
\newcommand{\He}{\ensuremath{{\what H}}}
\newcommand{\Ne}{\ensuremath{{\what N}}}
\newcommand{\Ns}{\ensuremath{\what N_{\star}}}

\newcommand{\dphi}{\ensuremath{\what{\delta\phi}}}
\newcommand{\dph}{\ensuremath{\delta\phi}}

\newcommand{\what}{\ensuremath{\widehat}}

\def\aal{{\bar\alpha}}
\def\bbet{{\bar\beta}}
\def\al{{\alpha}}

\def\bt{{\beta}}

\def\th{{\theta}}

\newcommand{\Trh}{\ensuremath{T_{\rm rh}}}
\newcommand{\sg}{\ensuremath{\phi}}

\newcommand{\sgx}{\ensuremath{\phi_\star}}
\newcommand{\sgf}{\ensuremath{\phi_{\rm f}}}
\newcommand{\xsg}{\ensuremath{x_{\phi}}}
\newcommand{\xst}{\ensuremath{x_{\star}}}

\newcommand{\ld}{\ensuremath{\lambda}}

\newcommand{\se}{\ensuremath{\widehat \phi}}
\newcommand{\sex}{\ensuremath{\widehat{\phi}_\star}}

\newcommand{\geu}{\ensuremath{\widehat g}}

\def\trns{transplanckian}
\def\Ka{K\"{a}hler potential}

\def\sub{subplanckian}

\def\FHI{IG inflation~}

\renewcommand{\arg}{\ensuremath{{\small\sf arg}}}

\newcommand{\bicep}{{B{\scshape icep}2}}
\newcommand{\plk}{\emph{Planck}}

\begin{document}

\thispagestyle{empty}

\title[]{\boldmath\Large\bfseries\scshape Reconciling
Induced-Gravity Inflation in Supergravity \\ With the
\emph{Planck} 2013 \& Bicep2 Results}

\author{\large\bfseries\scshape C. Pallis}
\address[] {\sl Departament de F\'isica Te\`orica and IFIC,\\
Universitat de Val\`encia-CSIC, \\ E-46100 Burjassot, SPAIN}

\begin{abstract}{{\bfseries\scshape Abstract} \\
\par We generalize the embedding of induced-gravity inflation
beyond the no-scale Supergravity presented in \cref{pallis}
employing two gauge singlet chiral superfields, a superpotential
uniquely determined by applying a continuous $R$ and a discrete
$\mathbb{Z}_n$ symmetries, and a logarithmic \Ka\ including all
the allowed terms up to fourth order in powers of the various
fields. We show that, increasing slightly the prefactor $(-3)$
encountered in the adopted \Ka, an efficient enhancement of the
resulting tensor-to-scalar ratio can be achieved rendering the
predictions of the model consistent with the recent \bicep\
results, even with \sub\ excursions of the original inflaton
field. The remaining inflationary observables can become
compatible with the data by mildly tuning the coefficient involved
in the fourth order term of the \Ka\ which mixes the inflaton with
the accompanying non-inflaton field. The inflaton mass is
predicted to be close to $10^{14}~\GeV$. }
\\ \\
{\ftn \sf Keywords: Cosmology, Supersymmetric models, Supergravity, Modified Gravity};\\
{\ftn \sf PACS codes: 98.80.Cq, 11.30.Qc, 12.60.Jv, 04.65.+e, 04.50.Kd}%
\\ \\
{\it Dedicated to the memory of my High School teacher, V.
Aspiotis.}\\ \\ \publishedin{{\sl  J. Cosmol. Astropart. Phys.}
{\bf 10}, {058} (2014)}
\end{abstract} \maketitle



\setcounter{page}{1} \pagestyle{fancyplain}

\rhead[\fancyplain{}{ \bf \thepage}]{\fancyplain{}{\sl Reconciling
IG Inflation in SUGRA With the {\it Planck} 2013 \& B{\ftn ICEP}2
Results}} \lhead[\fancyplain{}{ \sl \leftmark}]{\fancyplain{}{\bf
\thepage}} \cfoot{}

\newpage

\tableofcontents\vskip-1.3cm\noindent\rule\textwidth{.5pt}

\section{Introduction}\label{intro} %

Although compatible with the \plk\ (and {WMAP}) data \cite{plin},
the models of \emph{induced-gravity} ({\sf\ftn IG}) inflation
\cite{induced} formulated within standard \emph{Supergravity}
({\sf\ftn SUGRA}) yield \cite{pallis} a low tensor-to-scalar ratio
$r\simeq0.004$ which fails to approach the recent \bicep\ results
\cite{gws} -- for other recent incarnations of IG inflation see
\cref{gian,rena}. More specifically, the \bicep\ collaboration has
detected a B-mode in the polarization of the cosmic microwave
background radiation at large angular scales. If this observation
is attributed to the primordial gravity waves predicted by
inflation, it implies \cite{gws} $r=0.16^{+0.06}_{-0.05}$ -- after
substraction of a dust foreground. Despite the fact that this
result is subject to considerable uncertainties \cite{gws2, gws1}
and its interpretation as a detection of primordial gravitational
waves becomes more and more questionable \cite{liddle}, it
motivates us to explore how IG inflation can also accommodate
large $r$'s -- for similar recent attempts see
\cref{rStar,rEllis,aroest}. In particular, taking into account
both the \plk\ \cite{plin} and \bicep\ \cite{gws} data we find a
simultaneously compatible region \cite{rcom} \beq 0.06\lesssim
r\lesssim0.135 \label{rgw}\eeq at 95$\%$ \emph{confidence level}
(c.l.) which can be considered as the most exciting region where
$r$ values may be confined for models with low running, $\as$, of
the (scalar) spectral index, $\ns$.

In this paper we show that modifying modestly the implementation
of IG inflation beyond the no-scale SUGRA \cite{noscale} we can
ensure a sizable augmentation of the resulting $r$'s \emph{with
respect to} ({\sf\ftn w.r.t}) those obtained in the models
presented in \cref{pallis}. The key-ingredient of our
generalization is the variation of the numerical prefactor
encountered in the adopted \Ka. We show that increasing the
conventional value $(-3)$ of this prefactor by an amount of order
$0.01$, the inflationary potential acquires a moderate inclination
accommodating, thereby, observable $r$'s reconcilable with
\Eref{rgw}. In this set-up IG inflation, although less predictive
than its realization in no-scale SUGRA, preserves a number of
attractive features \cite{R2r, pallis}. Most notably, the super-
and \Ka s are fixed by an $R$ and a discrete $\mathbb{Z}_n$
symmetries, inflation is realized using \sub\ values of the
initial (non-canonically normalized) inflaton field, the radiative
corrections remain under control and the perturbative unitarity is
respected up to the reduced Planck scale,
$\mP=2.44\cdot10^{18}~\GeV$ \cite{riotto, R2r, pallis}.

Below we generalize in Sec.~\ref{fhim} the formulation of IG
inflationary models within SUGRA. In Sec.~\ref{fhi} we present the
basic ingredients of these models, derive the inflationary
observables and test them against observations. We end-up with a
brief analysis of the UV behavior of these models in
Sec.~\ref{fhi3} and the summary of our conclusions in
Sec.~\ref{con}. Throughout we follow closely the notation and the
conventions adopted in \cref{pallis}, whose Sections, Equations,
Tables and Figures are referred including a prefix ``R.1''. E.g.
\rEref{3.6} denotes Eq.~(3.6) of \cref{pallis}.

\section{Generalizing the Embedding of the IG Inflation in SUGRA}\label{fhim}

According to the scheme proposed in \cref{pallis}, the
implementation of IG inflation in SUGRA requires at least two
singlet superfields, i.e., $z^\al=\Phi, S$, with $\Phi$ ($\al=1$)
and $S$ ($\al=2)$ being the inflaton and a stabilized field
respectively. The superpotential $W$ of the model has the form
\beq\label{Whi} W= {\ld\mP^2\over\ck} S\lf
\fk-1/2\rg\>\>\>\mbox{with}\>\>\>\fk(\Phi)=\ck
\frac{\Phi^n}{\mP^n}+\sum_{k=1}^\infty\lambda_{k}\frac{\Phi^{2kn}}{\mP^{2kn}}\eeq
which is {\sf\ftn (i)} invariant under the action of a global
$\mathbb{Z}_n$ discrete symmetry, i.e., \beq W \to\ \,
W\>\>\>\mbox{for}\>\>\>\Phi\ \to\ -\Phi\,,\label{Zn} \eeq  and
{\sf\ftn (ii)} consistent with a continuous $R$ symmetry under
which \beq W \to\ e^{i\varphi}\, W\>\>\>\mbox{for}\>\>\>S\ \to\
e^{i\varphi}\,S\>\>\>\mbox{and}\>\>\>\fk\ \to\ \fk\,.\label{Rsym}
\eeq
Confining ourselves to $\Phi<\mP$ and assuming relatively low
$\lambda_{k}$'s we hereafter neglect the second term in the
definition of $\fk$ in \Eref{Whi}. As shown in \cref{pallis}, $W$
in \Eref{Whi} leads to a spontaneous breaking of $\mathbb{Z}_n$ at
the SUSY vacuum which lies at the direction
\beq \vev{S}=0\>\>\>\mbox{and}\>\>\> \vev{\fk}=1/2,\label{vevs}
\eeq
where we take into account that the phase of $\Phi$, $\arg\Phi$,
is stabilized to zero. If $\fk$ is the holomorphic part of the
frame function $\Omega$ and dominates it, \Eref{vevs} assures a
transition to the conventional Einstein gravity realizing,
thereby, the idea of IG \cite{induced}. Our main point in this
paper is that this construction remains possible for a broad class
of relations between $\Omega$ and the \Ka\ $K$.

Indeed, if we perform a conformal transformation defining the JF
metric $g_{\mu\nu}$ through the relation
\beq \label{weyl}
\geu_{\mu\nu}=-\frac{\Omega}{3(1+m)}g_{\mu\nu}~~\Rightarrow~~\left\{\bem
\sqrt{-\what{ \mathfrak{g}}}={\Omega^2\over
9(1+m)^2}\sqrt{-\mathfrak{g}}\>\>\>\mbox{and}\>\>\>
\geu^{\mu\nu}=-{3(1+m)\over\Omega}g^{\mu\nu}, \hfill \cr
\rce=-{3(1+m)\over\Omega}\left(\rcc-\Box\ln \Omega+3g^{\mu\nu}
\partial_\mu \Omega\partial_\nu \Omega/2\Omega^2\right) \hfill
\cr\eem
\right.\eeq
where $m$ is a dimensionless (small in our approach) parameter
which quantifies the deviation from the standard set-up
\cite{linde1}, the EF action
\beq\label{Saction1} {\sf S}=\int d^4x \sqrt{-\what{
\mathfrak{g}}}\lf-\frac{1}{2}\mP^2 \rce +K_{\al\bbet}\geu^{\mu\nu}
\partial_\mu z^\al \partial_\nu z^{*\bbet}-\Ve\rg, \eeq
-- where $\Ve$ is the F--term SUGRA scalar potential given below
--, is written in the JF as follows \cite{linde1}
\beq {\sf S}=\int d^4x \sqrt{-\mathfrak{g}}\lf\frac{\mP^2\Omega
\rcc}{6(1+m)}+\frac{\mP^2}{4(1+m)\Omega}\partial_\mu\Omega\partial^\mu\Omega
-\frac{1}{(1+m)}\Omega K_{\al{\bbet}}\partial_\mu z^\al
\partial^\mu z^{*\bbet}-V  \rg\label{action2}\eeq
with $V ={\Omega^2\over9(1+m)^2}\Ve$ being the JF potential. If we
specify the following relation between $\Omega$ and $K$,
\beq-\Omega/3(1+m)
=e^{-K/3(1+m)\mP^2}\>\Rightarrow\>K=-3(1+m)\mP^2\ln\lf-\Omega/3(1+m)\rg,\label{Omg1}\eeq
and employ the definition  \cite{linde1} of the purely bosonic
part of the on-shell value of the auxiliary field
\beq {\cal A}_\mu =i\lf K_\al\partial_\mu z^\al-K_\aal\partial_\mu
z^{*\aal}\rg/6, \label{Acal1}\eeq
we arrive at the following action
\beq {\sf S}=\int d^4x
\sqrt{-\mathfrak{g}}\lf\frac{\mP^2\Omega\rcc}{6(1+m)}+\mP^2\lf\Omega_{\al{\bbet}}-\frac{m\Omega_{\al}\Omega_{\bbet}}{(1+m)\Omega}\rg\partial_\mu
z^\al \partial^\mu z^{*\bbet}- \frac{\Omega{\cal A}_\mu{\cal
A}^\mu}{(1+m)^3\mP^2}-V \rg, \label{Sfinal}\eeq
where ${\cal A}_\mu$ in \Eref{Acal1} takes the form
\beq {\cal A}_\mu =-i(1+m)\mP^2\lf \Omega_\al\partial_\mu
z^\al-\Omega_\aal\partial_\mu z^{*\aal}\rg/2\Omega\,.
\label{Acal}\eeq
It is clear that \eqs{Sfinal}{Acal} reduce to \reqs{2.3}{2.4}
respectively for $m=0$. The choice $m\neq0$, although not
standard, is perfectly consistent with the idea of IG. Indeed, as
in \cref{pallis} we adopt the following form for the frame
function
\beqs\beq -\Omega/3(1+m)=\fk(\Phi)+{\fk}^*(\Phi^*)-\Fk\lf|\Phi|,
|S|\rg/3(1+m), \label{Omg}\eeq
where $\Fk$ includes the kinetic terms for the $z^\al$'s and takes
the form
\beq \label{Fkdef} \Fk\lf|\Phi|, |S|\rg=
{|S|^2+|\Phi|^2\over\mP^2}\,-\, {\kx|S|^4+2\kpp|\Phi|^4+2\ksp
|S|^2|\Phi|^2\over\mP^4},\eeq\eeqs
with sufficiently small coefficients $k_{\al\bt}$ i.e.
$k_{\al\bt}\ll\ck$. As a consequence, $\Fcr$ represents the
non-minimal coupling to gravity and so \Eref{vevs} dynamically
generates $\mP$. As for $m=0$, when the dynamics of the $z^\al$'s
is dominated only by the real moduli $|z^\al|$ or if $z^\al=0$ for
$\al\neq1$ \cite{linde1},  we can obtain ${\cal A}_\mu=0$ in
\Eref{Sfinal}. The only difference w.r.t the case with $m=0$ is
that now the scalar fields $z^\al$ have not canonical kinetic
terms in the JF due to the term proportional to
$\Omg_\al\Omg_\bbet\neq\delta_{\al\bbet}$. This fact does not
cause any problem, since the canonical normalization of the
inflaton keeps its strong dependence on $\ck$ included in $\fk$
whereas the non-inflaton fields become heavy enough during
inflation and so, they do not affect the dynamics -- see
\Sref{fgi1}. Note that our present set-up lies on beyond the
no-scale SUGRA embedding of IG inflation since the framework of
the no-scale SUGRA \cite{noscale} is defined by supplementing
\rEref{2.8} with the imposition $m=0$. Indeed, only under this
condition the cosmological constant term into the EF F--term SUGRA
scalar potential -- see below -- vanishes.

The resulting through \Eref{Omg1} \Ka\ is
\beq  \Khi=-3(1+m)\mP^2\ln\lf
\fk+\fk^*-{|S|^2+|\Phi|^2\over3(1+m)\mP^2}+{\kx|S|^4+2\kpp|\Phi|^4+2\ksp
|S|^2|\Phi|^2\over3(1+m)\mP^4}\rg.\label{Kolg}\eeq
Recall that the fourth order term for $S$ is included to cure the
problem of a tachyonic instability occurring along this direction
\cite{linde1} and the remaining terms of the same order are
considered for consistency -- the factors of $2$ are added just
for convenience. Alternative solutions to the aforementioned
problem of the tachyonic instability are recently identified in
\cref{ketov, nil, noK}.

\section{The Inflationary Scenario}\label{fhi}

In this section we describe the inflationary potential of our
model in Sec.~\ref{fgi1}. We then exhibit a number of constraints
imposed (\Sref{fhi2}) and present our analytic and numerical
results in Sec.~\ref{gan} and \ref{gnum} respectively.

\subsection{The Inflationary Potential}\label{fgi1}

The EF F--term (tree level) SUGRA scalar potential, $\Vhio$, of IG
inflation is obtained from $\Whi$ and $\Khi$ in \eqs{Whi}{Kolg}
respectively by applying (for $z^\al=\Phi,S$) the well-known
formula
\beqs\beq \Vhio=e^{\Khi/\mP^2}\left(K^{\al\bbet}{\rm F}_\al {\rm
F}^*_\bbet-3\frac{\vert W\vert^2}{\mP^2}\right)
=e^{K/\mP^2}K^{SS^*}\,
|W_{,S}|^2=\frac{\ld^2\mP^4|2\fk-1|^2}{4\ck^2\fsp\fr^{2+3m}}\,,\label{Vhig}\eeq
where ${\rm F}_\al=W_{,z^\al} +K_{,z^\al}W/\mP^2$ and $S$ is
placed at the origin. Here we take into account that
\beq
e^{K/\mP^2}=\fr^{-3(1+m)}\>\>\>\mbox{and}\>\>\>K^{SS^*}={\fr/\fsp}.\label{Vhigg}\eeq
The functions $\fr$ and $\fsp$, defined as follows -- cf.
\rEref{3.26}:
\bea \label{frsp} &&
\fr=-\frac{\Ohi}{3(1+m)}=2\ck{\xsg^n\over2^{n/2}}+{\xsg^2\over6(1+m)}+{\kpp\over12(1+m)}\xsg^4;\\
&& \label{frsp1}
\fsp=\mP^2\Omega_{,SS^*}=1-\ksp\xsg\>\>\>\mbox{with}\>\>\>\xsg={\sg/\mP},\eea\eeqs
are computed along the inflationary track, i.e., for
\beq \th=s=\bar s=0,\label{inftr} \eeq
using the standard parametrization for $\Phi$ and $S$
\beq \Phi=\:\frac{\phi}{\sqrt{2}}\,e^{i
\th/\mP}\>\>\>\mbox{and}\>\>\>S=\:\frac{s +i\bar
s}{\sqrt{2}}\,\cdot\label{cannor} \eeq
Given that $\fsp\ll\fr\simeq2\fk$ with $\ck\gg1$, $\Vhio$ in
\Eref{Vhig} is roughly proportional to $\xsg^{-3mn}$. Besides the
inflationary plateau which emerges for $m=0$ and studied in
\cref{pallis}, a chaotic-type potential (bounded from below) is
generated for $m<0$. More specifically, $\Vhio$ can be cast in the
following from -- cf. \srEref{3.25}{a}:
\beq\label{3Vhioo} \Vhio=\frac{\ld^2\mP^4\ft^2
\xsg^{-6m}\lf2^{1-n/2}\ck\xsg^{n-2}-f_{\phi\phi}/6(1+m)\rg^{-3m}}{4\ck^2\xsg^4\lf\ck
\xsg^{n-2}-2^{n/2-1} f_{\phi\phi}/6(1+m)\rg^2\fsp}, \eeq
where $f_{\phi\phi}=1-\kpp\xsg^2$ and $\ft=(2^{n/2-1}-\ck\xsg^n)$
coincides with $f_{\phi\phi}$ and $f_{\Phi}$ defined in
\rEref{3.10}. Confining ourselves to $n=2$ -- which, as we justify
in \Sref{gnum} consists the most interesting choice -- $\Vhio$
takes the form
\beq \Vhio=\frac{\ld^2\mP^4f_2^2\xsg^{-6m}}{4\ck^2\xsg^4\fsp}
\lf\ck-\frac{f_{\phi\phi}}{6(1+m)}\rg^{-(2+3m)} \simeq\frac{\ld^2
\mP^4\xsg^{-6m}}{4\fsp\ck^{2+3m}}\,,\label{3Vhiom}\eeq
whereas the corresponding EF Hubble parameter is
\beq \He_{\rm
IG}={\Vhio^{1/2}\over\sqrt{3}\mP}\simeq{\ld\mP\xsg^{-3m}\over2\sqrt{3\fsp}\ck^{1+3m/2}}\,\cdot
\label{He}\eeq

The stability of the configuration in \Eref{inftr} can be checked
verifying the validity of the conditions
\beq {\partial \Vhio/\partial\what\chi^\al}=0\>\>\>
\mbox{and}\>\>\>\what m^2_{
\chi^\al}>0\>\>\>\mbox{with}\>\>\>\chi^\al=\th,s,\bar
s,\label{Vcon} \eeq
where $\what m^2_{\chi^\al}$ are the eigenvalues of the mass
matrix with elements $\what
M^2_{\al\bt}={\partial^2\Vhio/\partial\what\chi^\al\partial\what\chi^\beta}$
and hat denotes the EF canonically normalized fields defined by
the kinetic terms in \Eref{Saction1} as follows
\beqs\beq \label{K3} K_{\al\bbet}\dot z^\al \dot
z^{*\bbet}=\frac12\lf\dot{\se}^{2}+\dot{\what
\th}^{2}\rg+\frac12\lf\dot{\what s}^2 +\dot{\what{\overline
s}}^2\rg,\eeq
where the dot denotes derivation w.r.t the JF cosmic time and the
hatted fields read
\beq  \label{Jg} {d\widehat \sg\over
d\sg}=\sqrt{K_{\Phi\Phi^*}}=J\simeq{\sqrt{6(1+m)}\over\xsg},\>\>\>
\what{\th}= J\,\th\xsg\>\>\>\mbox{and}\>\>\>(\what s,\what{\bar
s})=\sqrt{K_{SS^*}} {(s,\bar s)}.\eeq\eeqs
where $K_{SS^*}\simeq1/\ck\xsg^2$ -- cf. \eqs{Vhigg}{frsp1}. The
spinors $\psi_\Phi$ and $\psi_S$ associated with $S$ and $\Phi$
are normalized similarly, i.e.,
$\what\psi_{S}=\sqrt{K_{SS^*}}\psi_{S}$ and
$\what\psi_{\Phi}=\sqrt{K_{\Phi\Phi^*}}\psi_{\Phi}$. Integrating
the first equation in \Eref{Jg} we can identify the EF field:
\beq \se=\se_{\rm
c}+\sqrt{6(1+m)}\mP\ln\frac{\sg}{\vev{\sg}}\>\>\>\mbox{with}
\>\>\>\vev{\phi}=\frac{\sqrt{2}\mP}{\sqrt[n]{2\ck}},\label{se1}\eeq
where $\se_{\rm c}$ is a constant of integration and we take into
account \eqs{Whi}{vevs}.

Upon diagonalization of $\what M^2_{\al\bt}$, we construct the
mass spectrum of the theory along the path of \Eref{inftr}. Taking
advantage of the fact that $\ck\gg1$ and the limits $\kpp\to0$ and
$\ksp\to0$ we find the expressions of the relevant masses squared,
arranged in \Tref{tab4}, which approach rather well the quite
lengthy, exact expressions taking into account in our numerical
computation. In the limit $m=0$ the expressions in \rTref{1} are
recovered. We have numerically verified that the various masses
remain greater than $\Hhi$ during the last $50$ e-foldings of
inflation, and so any inflationary perturbations of the fields
other than the inflaton are safely eliminated. They enter a phase
of oscillations about zero with reducing amplitude and so the
$\xsg$ dependence in their normalization -- see \Eref{Jg} -- does
not affect their dynamics. As usually -- cf. \cref{pallis, R2r} --
the lighter eignestate of $\what M^2_{\al\bt}$ is $\what
m^2_{{s}}$ which here can become positive and heavy enough for
$\kx\gtrsim0.1$ -- see \Sref{gnum}.

\begin{table}[!t]
\renewcommand{\arraystretch}{1.4}
\bec\begin{tabular}{|c|c|l|}\hline
{\sc Fields} &{\sc Eingestates} & \hspace*{3.cm}{\sc Masses Squared}\\
\hline \hline
$1$ real scalar &$\what{\th}$ & $\what m^2_{\th}\simeq\ld^2
\mP^2\lf 2-2\ck\xsg^2f_2+3m f_2^2\rg$\\ && $ /6(1 + m) \ck^{4 +3 m}\xsg^{2(2 + 3 m)} \simeq4\He_{\rm IG}^2$\\
$2$ real scalars &$\what{s},~\what{\bar s}$ & $\what m^2_{
s}=\ld^2 \mP^2  \lf2 -6m - \ck \xsg^2+ 6\ks (1+m)f_2^2\rg$
\\&&$/6 (1+m)\ck^{3(1+m)}\xsg^{2(1+3m)}$\\ \hline
$2$ Weyl spinors & $\what{\psi}_\pm={\what{\psi}_{\Phi}\pm
\what{\psi}_{S}\over\sqrt{2}}$& $\what m^2_{ \psi\pm}\simeq\ld^2
\mP^2(2+3 m\fpp)^2/12(1 + m) \ck^{4 +3 m}\xsg^{2(2 + 3 m)}$
\\ \hline
\end{tabular}\ec
\renewcommand{\arraystretch}{1.}
\hfill \vchcaption[]{\sl\small The mass spectrum along the
inflationary trajectory in \Eref{inftr} for $m<0$ and $n=2$ in
\eqs{Whi}{Kolg}.}\label{tab4}
\end{table}

Inserting, finally, the mass spectrum of the model in the
well-known Coleman-Weinberg formula, we calculate the one-loop
corrected $\Vhi$
\beq\Vhi=\Vhio+{1\over64\pi^2}\lf \widehat m_{ \th}^4\ln{\widehat
m_{\th}^2\over\Lambda^2} +2 \widehat m_{ s}^4\ln{\widehat
m_{s}^2\over\Lambda^2}-4\widehat
m_{\psi_{\pm}}^4\ln{m_{\widehat\psi_{\pm}}^2\over\Lambda^2}\rg
,\label{Vhic}\eeq
where $\Lambda$ is a \emph{renormalization group} ({\sf\small RG})
mass scale. We determine it by requiring \cite{Qenq} $\Delta
V(\sgx)=0$ with $\Delta V=\Vhi-\Vhio$. To reduce the possible
\cite{Qenq} dependence of our results on the choice of $\Lambda$,
we confine ourselves to $\ld$ and $\kx$ values which do not
enhance these corrections -- see \Sref{gnum}.

\subsection{The Inflationary Requirements}\label{fhi2}

Based on $\Ve_{\rm IG}$ in \Eref{Vhic} we can proceed to the
analysis of \FHI in the EF \cite{induced}, employing the standard
slow-roll approximation \cite{review}. We have just to convert the
derivations and integrations w.r.t $\se$ to the corresponding ones
w.r.t $\sg$ keeping in mind the dependence of $\se$ on $\sg$,
\Eref{Jg}. In particular, the observational requirements which are
imposed on our inflationary scenario are outlined in the
following.

\paragraph{3.2.1} The number of e-foldings, $\Ns$, that
the scale $k_\star=0.05/{\rm Mpc}$ suffers during IG inflation has
to be adequate to resolve the horizon and flatness problems of
standard big bang, i.e., \cite{plin,nmi}
\begin{equation}
\label{Nhi}  \Ns=\int_{\se_{\rm f}}^{\se_{\star}}\,
\frac{d\se}{m^2_{\rm P}}\: \frac{\Ve_{\rm IG}}{\Ve_{\rm IG,\se}}
\simeq19.4+2\ln{\Ve_{\rm IG}(\sg_{\star})^{1/4}\over{1~{\rm
GeV}}}-{4\over 3}\ln{\Ve_{\rm IG}(\sg_{\rm f})^{1/4}\over{1~{\rm
GeV}}}+ {1\over3}\ln {T_{\rm rh}\over{1~{\rm
GeV}}}+{1\over2}\ln{\fr(\sg_{\rm f})\over \fr(\sg_\star)^{1/3}},
\end{equation}
where $\sg_\star~[\se_\star]$ is the value of $\sg~[\se]$ when
$k_\star$ crosses outside the inflationary horizon and $\sg_{\rm
f}~[\se_{\rm f}]$ is the value of $\sg~[\se]$ at the end of IG
inflation, which can be found from the condition
\beq \label{sr} {\sf max}\{\widehat\epsilon(\sg_{\rm
f}),|\widehat\eta(\sg_{\rm
f})|\}=1,~~~\mbox{where}\>\>\>\widehat\epsilon=
{\mP^2\over2}\left(\frac{\Ve_{\rm IG,\se}}{\Ve_{\rm
IG}}\right)^2\>\>\>\mbox{and}\>\>\>\widehat\eta= m^2_{\rm
P}~\frac{\Ve_{\rm IG,\se\se}}{\Ve_{\rm IG}} \eeq
are the well-known slow-roll parameters and $\Trh$ is the reheat
temperature after IG inflation, which is taken $\Trh=10^9~\GeV$
throughout.

\paragraph{3.2.2} The amplitude $A_{\rm s}$ of the power spectrum of the curvature perturbation
generated by $\sg$ at  $k_{\star}$ has to be consistent with
data~\cite{plin}
\begin{equation}  \label{Prob}
\sqrt{A_{\rm s}}=\: \frac{1}{2\sqrt{3}\, \pi\mP^3} \;
\frac{\Ve_{\rm IG}(\sex)^{3/2}}{|\Ve_{\rm
IG,\se}(\sex)|}={1\over2\pi\mP^2}\,\sqrt{\frac{\Vhi(\sgx)}{6\what\epsilon_\star}}
\simeq4.685\cdot 10^{-5},
\end{equation}
where the variables with subscript $\star$ are evaluated at
$\sg=\sg_{\star}$

\paragraph{3.2.3}  The remaining inflationary observables $\ns,\as$ and $r$ -- estimated through the relations:
\beq\label{ns} \mbox{\ftn\sf (a)}\>\>\ns=\:
1-6\widehat\epsilon_\star\ +\
2\widehat\eta_\star,\>\>\>\mbox{\ftn\sf (b)}\>\> \as
=\:2\left(4\widehat\eta_\star^2-(n_{\rm
s}-1)^2\right)/3-2\widehat\xi_\star\>\>\>
\mbox{and}\>\>\>\mbox{\ftn\sf (c)}\>\>r=16\widehat\epsilon_\star,
\eeq
with $\widehat\xi=\mP^4 {\Ve_{\rm IG,\widehat\sg} \Ve_{\rm
IG,\widehat\sg\widehat\sg\widehat\sg}/\Ve_{\rm IG}^2}$ -- have to
be consistent with the data \cite{plin},  i.e.,
\begin{equation}  \label{nswmap}
\mbox{\ftn\sf (a)}\>\>\ns=0.96\pm0.014,\>\>\>\mbox{\ftn\sf
(b)}\>\>-0.0314\leq a_{\rm s}\leq0.0046
\>\>\>\mbox{and}\>\>\>\mbox{\ftn\sf
(c)}\>\>r\leq0.135\>\>\>\mbox{at 95$\%$ c.l.},
\end{equation}
pertaining to the $\Lambda$CDM framework. The last inequality can
be complimented by the \bicep\ data as shown in \Eref{rgw}.

\paragraph{3.2.4.} Since SUGRA is an effective theory below $\mP$ the existence of higher-order terms in
$W$ and $K$, \eqs{Whi}{Kolg}, appears to be unavoidable. Therefore
the stability of our inflationary solutions can be assured if we
entail
\beq \label{subP}\mbox{\ftn\sf (a)}\>\>
\Vhi(\sg_\star)^{1/4}\leq\mP \>\>\>\mbox{and}\>\>\>\mbox{\ftn\sf
(b)}\>\>\sg_\star\leq\mP,\eeq
where $\mP$ is the UV cutoff scale of the effective theory for the
present models, as shown in \Sref{fhi3}.

\subsection{Analytic Results}\label{gan}

Plugging \eqs{3Vhiom}{Jg} into \Eref{sr} and taking $\kpp\simeq0$,
we obtain the following approximate expressions for the slow-roll
parameters
\bea \nonumber &\what\epsilon&=\frac{(2 + 3 m - 3m\ck \xsg^2 + (1
+ 3 m)\ksp \ck\xsg^4)^2}{3 (1 + m) \fsp^2
\fpp^2}\>\>\>\mbox{and}\>\>\>\what\eta=\frac{1}{3 (1 + m) \fsp^2
\fpp^2}\times\\&\times& 2 \Big[\xsg^2\Big(\ksp \lf\xsg^2 \lf 6 \ck
+ \ck^2\xsg^2 + \ksp \ck^2 \xsg^4\rg-11\rg-2 \ck \Big)+ 9 m^2
\fsp^2 \fpp^2\nonumber\\&+ & 4  + 6 m \fsp \fpp \lf2 + \ksp \xsg^2
(\ck \xsg^2-3)\rg\Big] \,.\label{gmsr1}\eea
Taking the limit of the expressions above for $\ksp\simeq0$ we can
analytically solve the condition in \Eref{sr} w.r.t $\xsg$. The
results are
\beq {\sg_{1\rm
f}\over\mP}=\sqrt{\frac{3(1-m)+2\sqrt{3(1+m)}}{3(1+m)\ck}}\>\>\>\mbox{and}\>\>\>{\sg_{2\rm
f}\over\mP}=\sqrt{\frac{1-9m+\sqrt{16+21m(3m-1)}}{3(1+m)\ck}}\,\cdot\label{sgf}\eeq
The end of IG inflation mostly occurs at $\sgf=\sg_{1\rm f}$
because this is mainly the maximal value of the two solutions
above.

Since $\sgf\ll\sgx$, we can estimate $\Ns$ through \Eref{Nhi}
neglecting $\sgf$. Our result is
\beqs\beq \Ns \simeq  (1 + m) \frac{3 m \ln\xst + \ln\lf 2 + 3 m -
3 \ck m \xst^2\rg}{|m|(2 + 3
m)}\>\>\>\mbox{with}\>\>\>\xst=\sgx/\mP \cdot\label{Ngm}\eeq
Ignoring the first term in the last equality and solving w.r.t
$\xst$ we extract $\sgx$ as follows
\beq\label{sm*}\sgx\simeq\mP/\sqrt{3|m|
\ck\re}\>\>\>\mbox{with}\>\>\>\re = e^{m (2 + 3 m)\Ns/(1 + m)}\eeq
Although a radically different dependence of $\sgx$ on $\Ns$
arises compared to the models of \cref{pallis} -- cf.
\srEref{3.17}{a} -- $\sgx$ can again remain \sub\ for large
$\ck$'s. Indeed,
\beq \label{fmsub} \sgx\leq\mP\>\>\>\Rightarrow\>\>\>\ck\geq
1/3|m|\re\,.\eeq\eeqs
As emphasized in \cref{pallis}, this achievement is crucial for
the viability of our proposal, since it protects the inflationary
computation against higher-order corrections from
non-renormalizable terms in $\fk$ -- see \Eref{Whi}. Note that
$\fk$ is totally defined in terms of $\Phi$. In other words, our
setting is independent of $\se_\star$ which can be found by
\Eref{se1} and remains \trns. Indeed, plugging \Eref{sm*} into
\Eref{se1} we find
\beq \se_\star\simeq\se_{\rm c}-\mP\sqrt{3(1 + m)/2}\lf \ln 3|m| +
m(2 + 3m)\Ns/(1 + m)\rg, \label{sme*}\eeq
which yields $\se_\star\simeq(8.9-12)\mP$ for $\se_{\rm c}=0$ and
$m=-(0.04-0.0625)$. Interestingly enough, $\se_\star$ turns out to
be independent of $\ck$ -- as the result shown in \rEref{3.18}.
Note that the independence of $\ksp$ is artificial since we ignore
$\ksp$ in the estimations below \Eref{sgf}.

Upon substitution of \Eref{sm*} into \Eref{Prob} we end up with
\begin{equation} \As^{1/2}\simeq\frac{\ld \ck^{-3 m/2}\xst^{2-3 m}}{4 \sqrt{2} \pi (2 - 3 \ck m \xst^2 + \ksp
\ck \xst^4)} \>\>\Rightarrow\>\>\ld \simeq \frac{4\pi\sqrt{2\As}
\lf\ksp + 9 \ck m^2\re(1 + 2\re)\rg}{3^{1+3 m/2}(|m|\re)^{1 +
3m/2}}\cdot \label{langm} \eeq
We remark that $\ld$ remains proportional to $\ck$ as for the
other models of \cref{pallis} -- cf.  \reqs{3.19}{3.29} -- but it
depends also on both $\ksp$ and $m$. Inserting \Eref{sm*} into
\Eref{gmsr1}, employing then \sEref{ns}{a} and expanding for
$\ck\gg1$ we find
\beqs\beq \label{nsgm} \ns\simeq 1 - 2 m\frac{ 3 m (1 + 4 \re) - 4
\re (1 - 3 m \re)}{1 + m} - 4 \ksp\frac{ 1 + 3 m (1 + 10 \re) - 36
m^2 \re}{9\ck m (1 + m) \re}\cdot \eeq
%
From this expression we see that $m<0$ and $\ksp<0$ assist us to
reduce $\ns$ sizably lower than unity as required in
\sEref{nswmap}{a}. Making use of Eqs.~(\ref{sm*}), (\ref{gmsr1})
and (\ref{ns}{\sf\ftn c}) we arrive at
\beq \label{ragm} r\simeq\frac{48 m^2 (1 + 2 \re)^2}{(1 + m) (1 +
3 m \re)^2} + \frac{32\ksp(1 + 2 \re)(1 - 6 m \re)}{3\ck\re (1 +
m) (1 + 3 m \re)^2}\cdot\eeq\eeqs
%
From the last result we conclude that primarily $|m|\neq0$ and
secondary $m<0$ help us to increase $r$.

\begin{table}[!t]
\begin{center}
\renewcommand{\arraystretch}{1.3}
{\small \begin{tabular}{|l||ll|ll|ll||ll|ll|ll|} \hline
\multicolumn{13}{|c|}{\sc Input Parameters}\\\hline\hline
$-m/10^{-2}$ &\multicolumn{2}{c|}{$4$}&\multicolumn{2}{c|}{$5$}&\multicolumn{2}{c||}{$6$}&\multicolumn{2}{c|}{$4$}&\multicolumn{2}{c|}{$5$}&\multicolumn{2}{c|}{$6$} \\
$\ck/10^{3}$ &  $1.9$&$5.3$&$5.59$ & $9$&$17.7$&$35.5$&$1.9$&$5.3$&$5.59$ & $9$&$17.7$&$35.5$\\
$-\ksp/10^{-2}$&$1.1$& $3$&$2$& $3.2$&$2.9$&$6$&$1.1$& $3$&$2$&
$3.2$&$2.9$&$6$
\\ \hline\hline
\multicolumn{13}{|c|}{\sc Output Parameters}\\\hline
&\multicolumn{6}{|c||}{\sc Analytic
Results}&\multicolumn{6}{|c|}{\sc Numerical Results}\\\cline{2-13}
\hline\hline
$\ld/0.1$ & $1.3$&$3.5$&$4$& $6.5$&$1.3$&$26$ &$1.1$&$3.2$& $3.3$&$5.3$&$9.7$&$19$\\
$\sgx/\mP$ &$0.57$ &$0.34$&$0.5$& $0.4$&$0.43$&$0.3$&$1$ &$0.6$ & $1$ &$0.8$&$1$&$0.7$\\
$\se_\star/\mP$ &\multicolumn{2}{c|}{$7.7$}&\multicolumn{2}{c|}{$8.65$}&\multicolumn{2}{c||}{$9.62$}&\multicolumn{2}{c|}{$9$}&\multicolumn{2}{c|}{$10.3$}&\multicolumn{2}{c|}{$11.6$}\\
$\sgf/0.01\mP$ &$3.4$ &$2$&$2$& $1.6$&$1$&$0.8$&$3.4$&$2$& $2$&$1.6$&$1$&$0.8$\\
$\Ns$ &  $56$&$57$&$56$& $56.5$&$56.3$&$56.8$&$55.5$&$55.6$& $56.4$&$56.7$&$56.9$&$56.9$\\
\hline
$\ns$
&\multicolumn{2}{c|}{$0.98$}&\multicolumn{2}{c|}{$0.976$}&\multicolumn{2}{c||}{$0.97$}&\multicolumn{2}{c|}{$0.975$}&\multicolumn{2}{c|}{$0.96$}&
\multicolumn{2}{c|}{$0.946$}\\
$r$
&\multicolumn{2}{c|}{$0.08$}&\multicolumn{2}{c|}{$0.12$}&\multicolumn{2}{c||}{$0.17$}&\multicolumn{2}{c|}{$0.07$}&\multicolumn{2}{c|}{$0.09$}&
\multicolumn{2}{c|}{$0.13$}\\
\hline
$\what m^2_\th/\Hhi^2\lf\sgx\rg$ &  $3.92$&$3.96$&$3.97$&$3.97$&\multicolumn{2}{c||}{$3.98$}&\multicolumn{2}{c|}{$3.9$}&\multicolumn{2}{c|}{$3.9$}&\multicolumn{2}{c|}{$3.9$}\\
$\what m^2_s/\Hhi^2\lf\sgf\rg$ &  $2.64$&$2.64$&$2.75$&$2.75$&\multicolumn{2}{c||}{$2.9$}&\multicolumn{2}{c|}{$3.64$}&\multicolumn{2}{c|}{$3.65$}&\multicolumn{2}{c|}{$3.67$}\\
\hline
\end{tabular}
}\end{center} \hfill \vchcaption[]{\sl\small Comparison between
the analytic and numerical results for six different sets of input
and output parameters of our model. We take $\kx=0.1$, $\kpp=0.5$
and $\Trh=10^{9}~\GeV$. Our numerical results are consistent with
Eqs.~(\ref{rgw}), (\ref{Nhi}), (\ref{Prob}), (\ref{nswmap}) and
(\ref{subP}).}\label{tab3}
\end{table}

To appreciate the validity of our analytic estimates, we test them
against our numerical ones. The relevant results are displayed in
Table~\ref{tab3}. We use six sets of input parameters -- see also
\Sref{gnum} -- and we present their response by applying the
formulae of \Sref{gan} (first six columns to the right of the
leftmost one) or using the formulae of \Sref{fhi2} with $\Vhi$
given in \Eref{Vhic} (next six columns). We see that the results
are quite close to each other with an exception regarding $\sgx$
whose the numerical and analytic values appreciably differ. This
fact can be attributed to the inaccuracy of \Eref{sm*} whose the
derivation is based on a number of efficient simplifications.
Despite this deviation, the absence of $\sgf$ from \Eref{Ngm}
assists us to evaluate rather accurately $\Ns$ and the analytic
values of $\se_\star$, $r$ and $\ns$ are rather close to the
numerical ones. As anticipated in \Eref{sme*}, $\se_\star$ is
independent of $\ck$ (and $\ksp$). Finally, from the two last rows
of \Tref{tab3} we see that the formulas of \Tref{tab4} are
reliable enough. As can be deduced by the relevant expressions,
$\what m^2_s$ is a monotonically increasing function of $\xsg$ and
so its minimal value is encountered for $\sg=\sgf$. On the
contrary, the minimal $\what m^2_\th$ is located at $\sg=\sgx$.

It is clear that the $\ns$ and $r$ values obtained in \Tref{tab3}
are perfectly consistent with both the \plk\ and \bicep\ results
-- cf. Eqs.~(\ref{rgw}) and (\ref{ns}{\sf\ftn a,b}). Furthermore,
the resulting $r$ remains constant for constant $m$ and $\ns$ and
is independent on $\ld$ (or $\ck$). This feature is verified by
our analytical estimate in \Eref{ragm} from which we observe that
the dominant contribution originates from the first fraction,
which is independent of $\ksp$ and $\ck$, whereas the correction
of the second fraction is suppressed by the inverse power of
$\ck$. The most impressive point, however, is that these large $r$
values are accommodated with \sub\ values of $\phi$. As first
stressed in \cref{nmi}, this fact does not contradict to the Lyth
bound \cite{lyth}, since the latter bound is applied to the EF
canonically normalized inflaton field $\what \phi$ which remains
\trns\ and close to the value shown in \Eref{sme*}. Therefore,
large $r$'s do not necessarily \cite{rRiotto} correlate with
\trns\ excursions of $\sg$ within IG inflation.

\subsection{Numerical Results} \label{gnum}

As shown in \eqss{Whi}{Kolg}{Nhi}, this inflationary scenario
depends on the parameters:
$$\ld,\>\ck,\>m,\>\kx,\>\ksp,\>\kpp\>\>\>\mbox{and}\>\>\>\Trh.$$
Besides the free parameters employed in \rSref{3.3.3}, we here
have $m$ which is constrained to negative values in order to
ensure the boundedness from below of $\Vhio$ -- see \Eref{3Vhiom}.
Using the reasoning of \rSref{3.3.3}, we set $\kpp=0.5$ and
$\Trh=10^9~\GeV$. On the other hand, $\what m_{s}^2$ becomes
positive with $\ks$'s lower than those used in \rSref{3.3.3} since
positive contributions from $m<0$ arise here -- see in
\Tref{tab4}. Moreover, due to the relatively large $\ld$'s
encountered in our scheme, if $\ks$ takes a value of order unity
$\what m_{s}^2$ grows more efficiently than in the cases with
$m=0$, rendering thereby the radiative corrections in \Eref{Vhic}
sizeable for very large $\ck$'s. To avoid such a certainly
unpleasant dependence of the model predictions on the radiative
corrections we tune somehow $\kx$ to lower values than those used
in \rSref{3.3.3}. E.g. we set $\kx=0.1$ throughout. For the same
reason we confine ourselves to the lowest possible $n$, $n=2$.
\eqss{Nhi}{Prob}{subP} assist us to restrict $\ld$ (or $\ck\geq1$)
and $\sgx$. By adjusting $m$ and $\ksp$ we can achieve not only
$\ns$'s in the range of \sEref{nswmap}{a} but also $r$'s in the
optimistic region of \Eref{rgw}.

The structure of $\Vhi$ as a function of $\sg$ for $m<0$ (and
$n=2$) is visualized in \Fref{fig3m}, where we depict $\Vhi$
versus $\sg$ for $\sgx=\mP$ and the selected values of $\ld, \ksp$
and $m$, shown in the label. These choices require that $\ck$'s
are $(1.7, 5.6, 26)\cdot10^3$ and result to $\ns=0.96$ and
$r=0.053, 0.096, 0.16$ for increasing $|m|$'s -- light gray, black
and gray line correspondingly. It would be instructive to compare
\Fref{fig3m} with \rFref{1}, where $\Vhi$ for $m=0$ is displayed
-- the fact that we employ a vanishing $\ksp$ in \rFref{1} does
not invalidate the comparison since the impact of $\ksp$ on the
form of $\Vhi$ is almost invisible. We remark that in \Fref{fig3m}

\begin{figure}[!t]\vspace*{-.44in}\begin{tabular}[!h]{cc}\begin{minipage}[t]{7.in}
\hspace*{.6in}
\epsfig{file=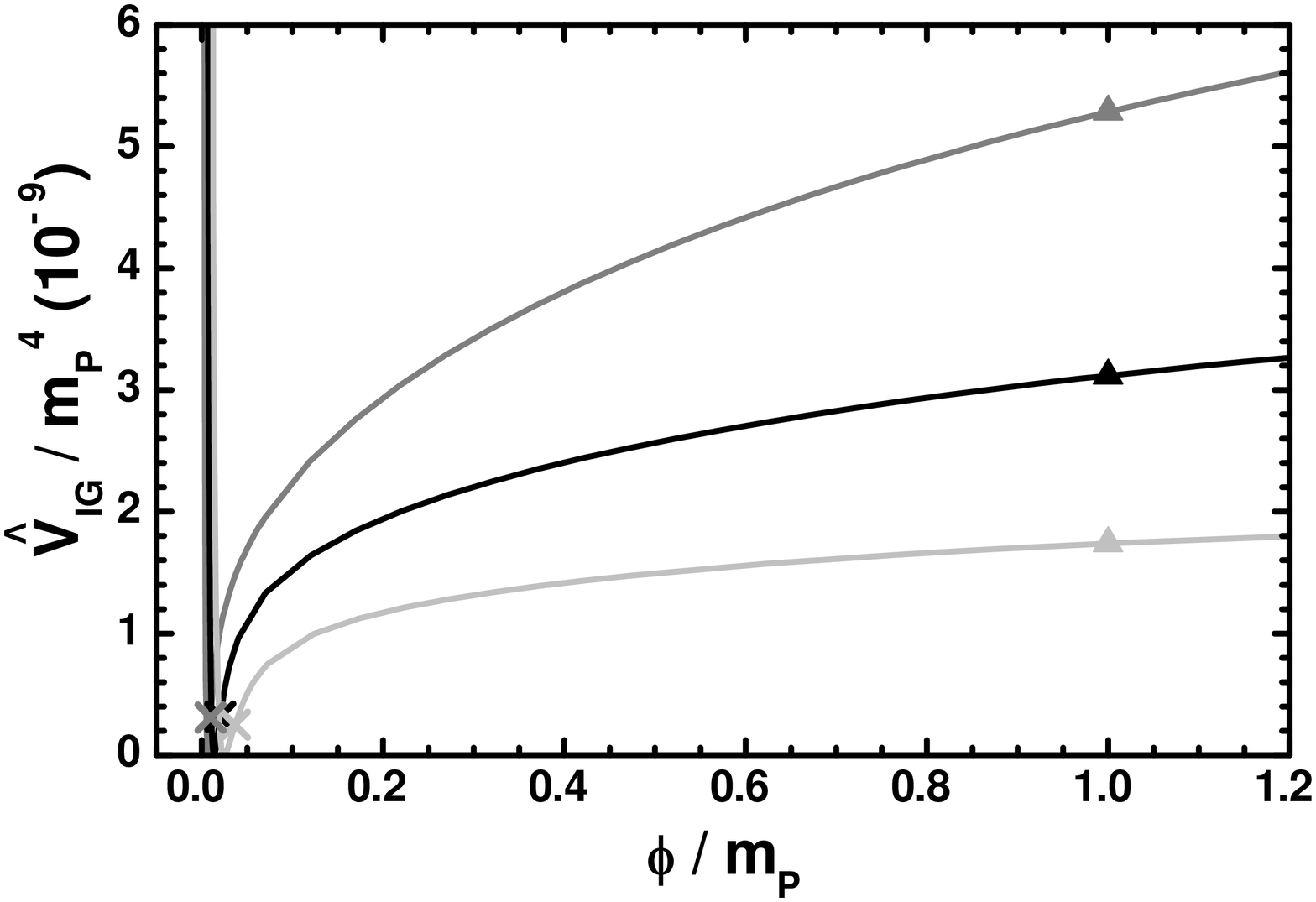,height=3.65in,angle=-90}\end{minipage}
&\begin{minipage}[h]{3.in}
\hspace{-3.5in}{\vspace*{-2.5in}\includegraphics[height=7.9cm,angle=-90]
{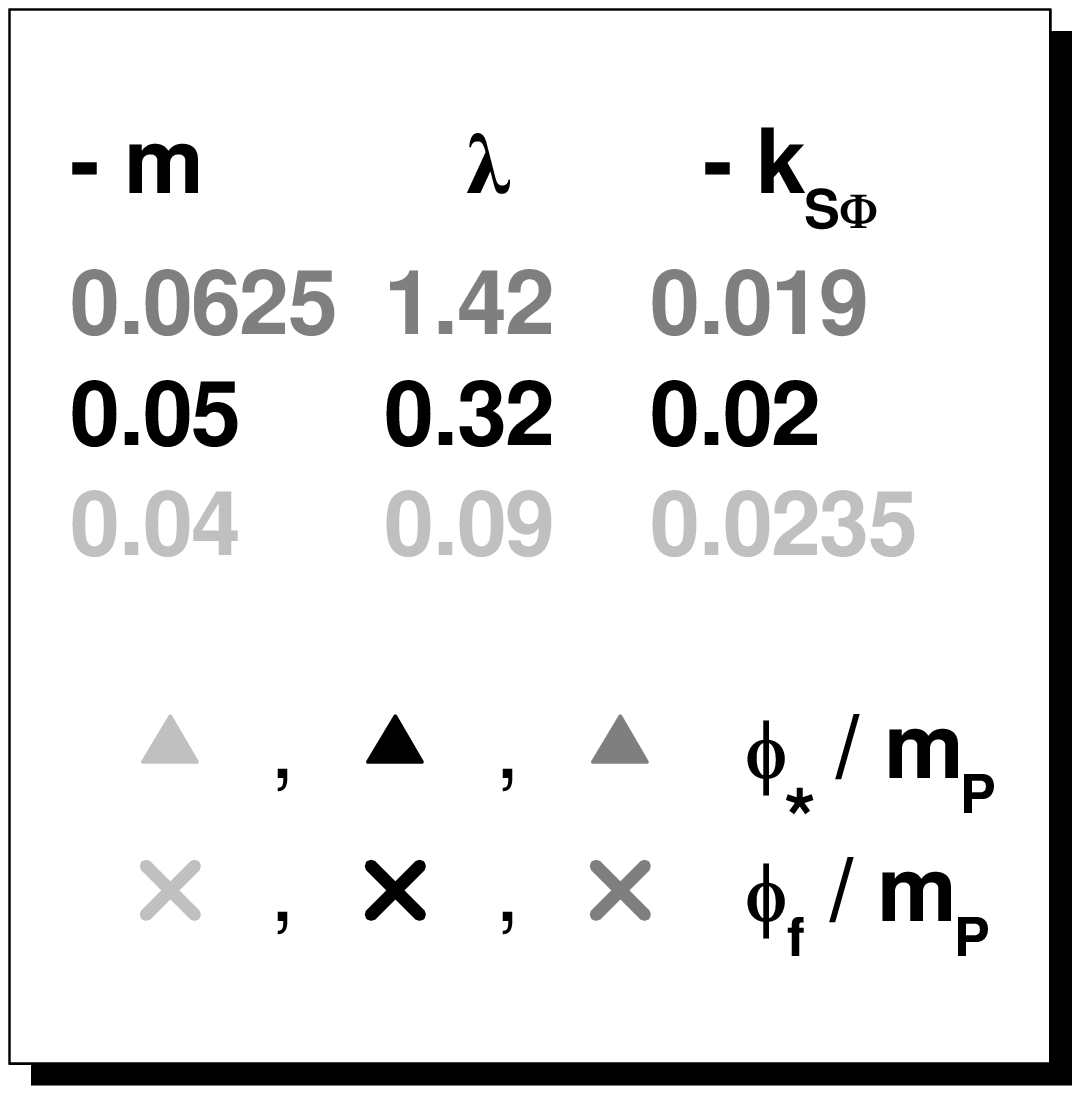}}\end{minipage}
\end{tabular}  \hfill \vchcaption[]{\sl \small The inflationary
potential $\Vhi$ as a function of $\sg$ for $n=2$. The light gray,
black and gray line is obtained by setting $-m=0.04,0.05,0.0625$,
$\ld=0.09,0.32,1.42$ and $-\ksp\simeq0.0235,0.02,0.019$
respectively. The values corresponding to $\sgx$ and $\sgf$ are
also depicted.}\label{fig3m}
\end{figure}

\begin{itemize}

\item[{\sf\ftn (i)}] The values of $\Vhio$ for $\sg=\sgx$ are one
order of magnitude larger than those encountered in \rFref{1};
actually $\Vhio^{1/4}$ approaches the SUSY grand-unification
scale, $2\cdot10^{16}~\GeV$, which is imperative -- see, e.g.,
\cref{rRiotto} -- for achieving $r$ values of order $0.1$;

\item[{\sf\ftn (ii)}] $\Vhio$ close to $\sg=\sgx$ acquires a
steeper slope which increases with $|m|$ and results to an
enhancement of $\what \epsilon_\star$ -- see \Eref{gmsr1} -- and,
via \sEref{ns}{c}, of $r$.

\end{itemize}
Another difference of the present set-up regarding those of
\cref{pallis} is that for $m=0$ we obtain constantly
$\eta_\star<0$ whereas we here obtain $\eta_\star>0$ for
$\ns>0.97$ and $\eta_\star<0$ for lower $\ns$ values.

\begin{figure}[!t]\vspace*{-.12in}
\hspace*{-.19in}
\begin{minipage}{8in}
\epsfig{file=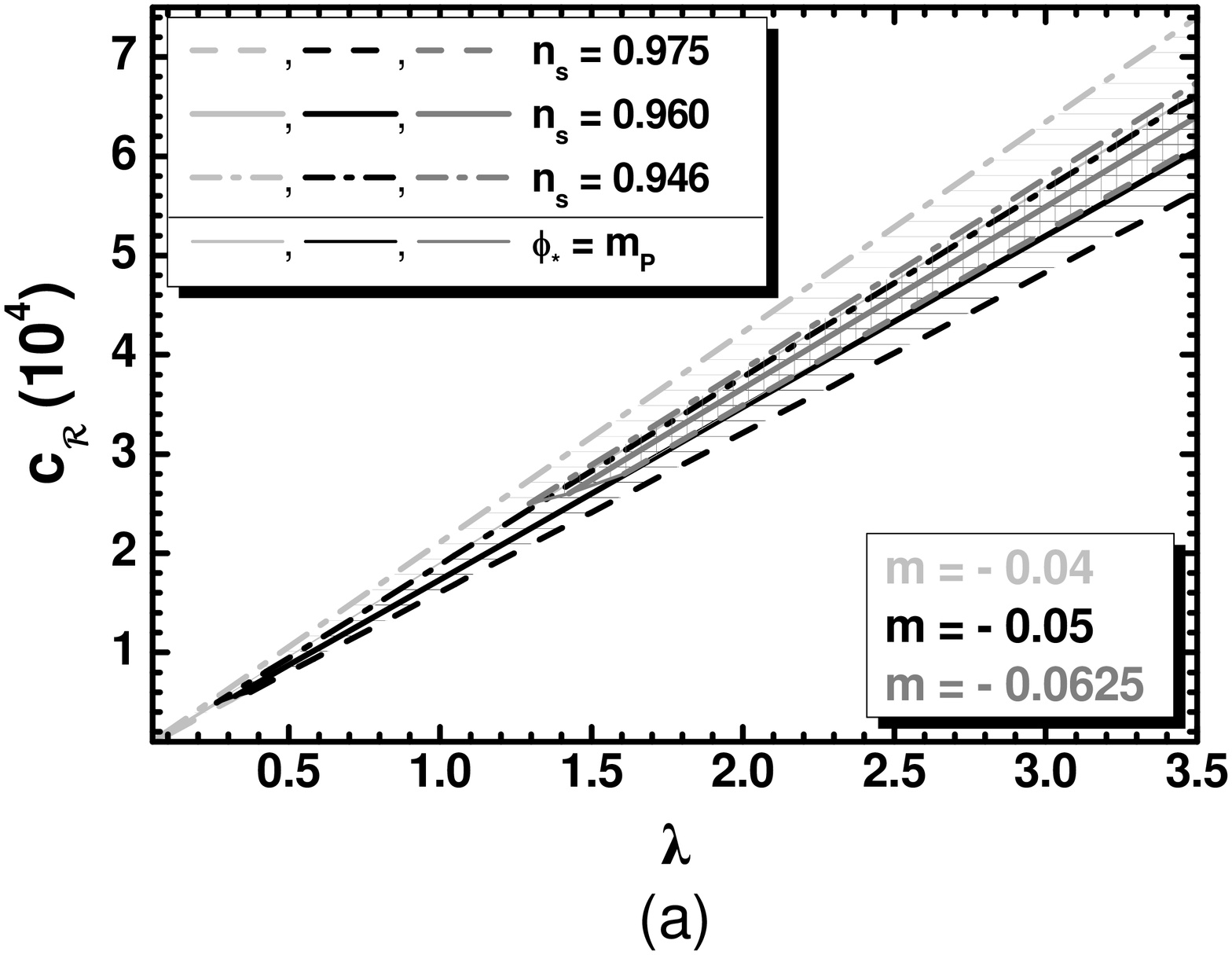,height=3.6in,angle=-90}
\hspace*{-1.2cm}
\epsfig{file=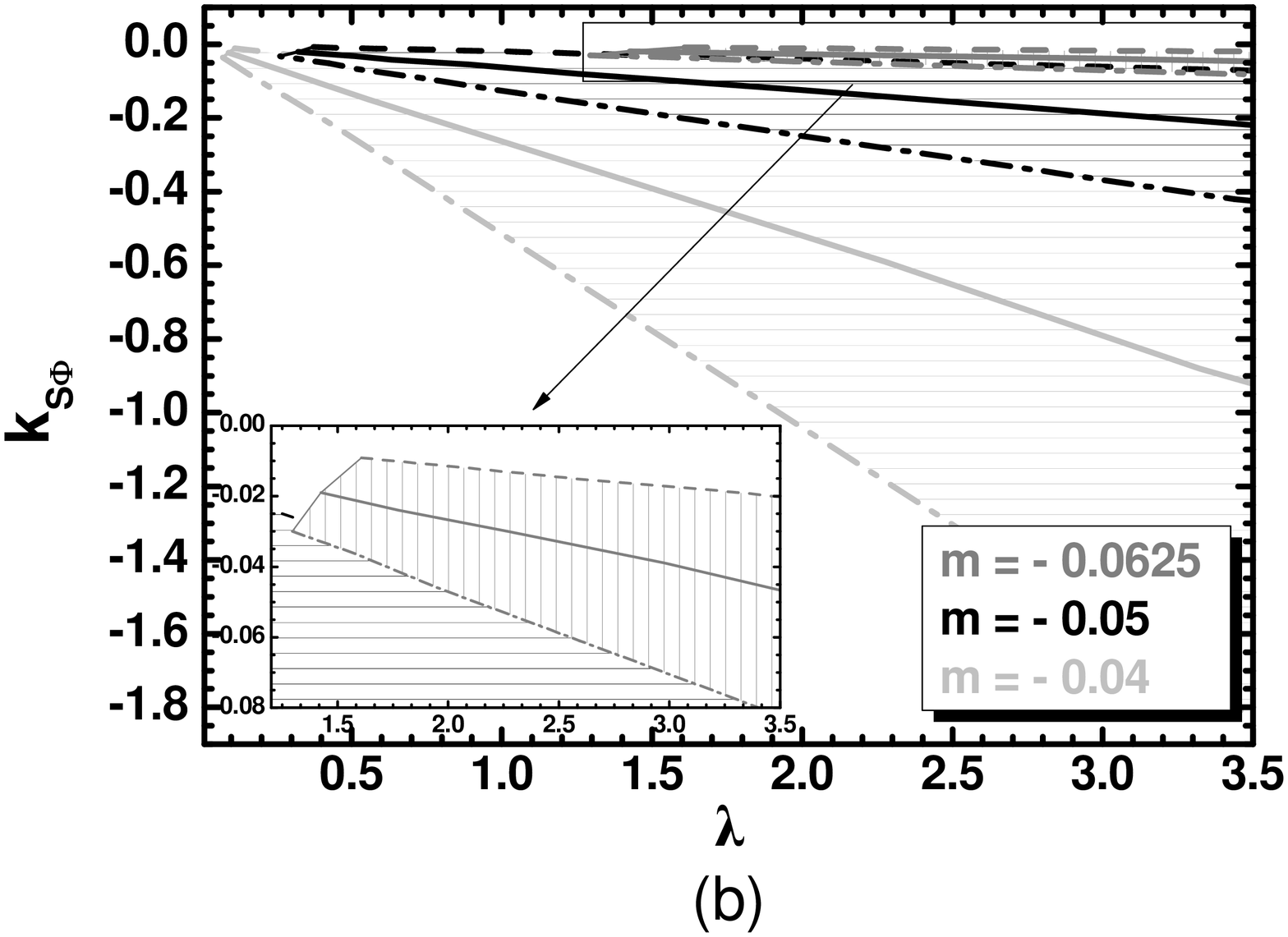,height=3.6in,angle=-90} \hfill
\end{minipage}
\hfill \hspace*{-.19in}
\begin{minipage}{8in}
\epsfig{file=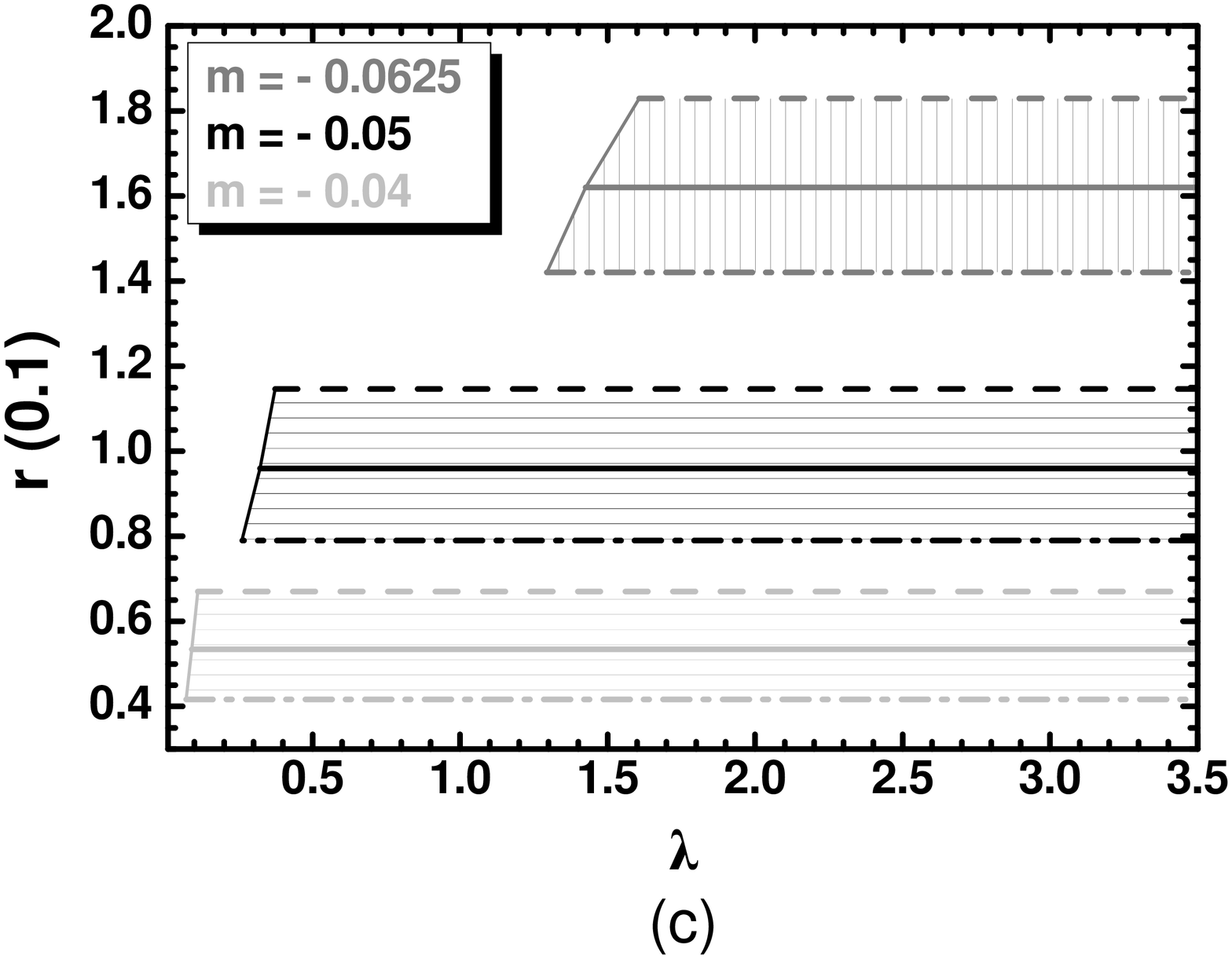,height=3.6in,angle=-90}
\hspace*{-1.2cm}
\epsfig{file=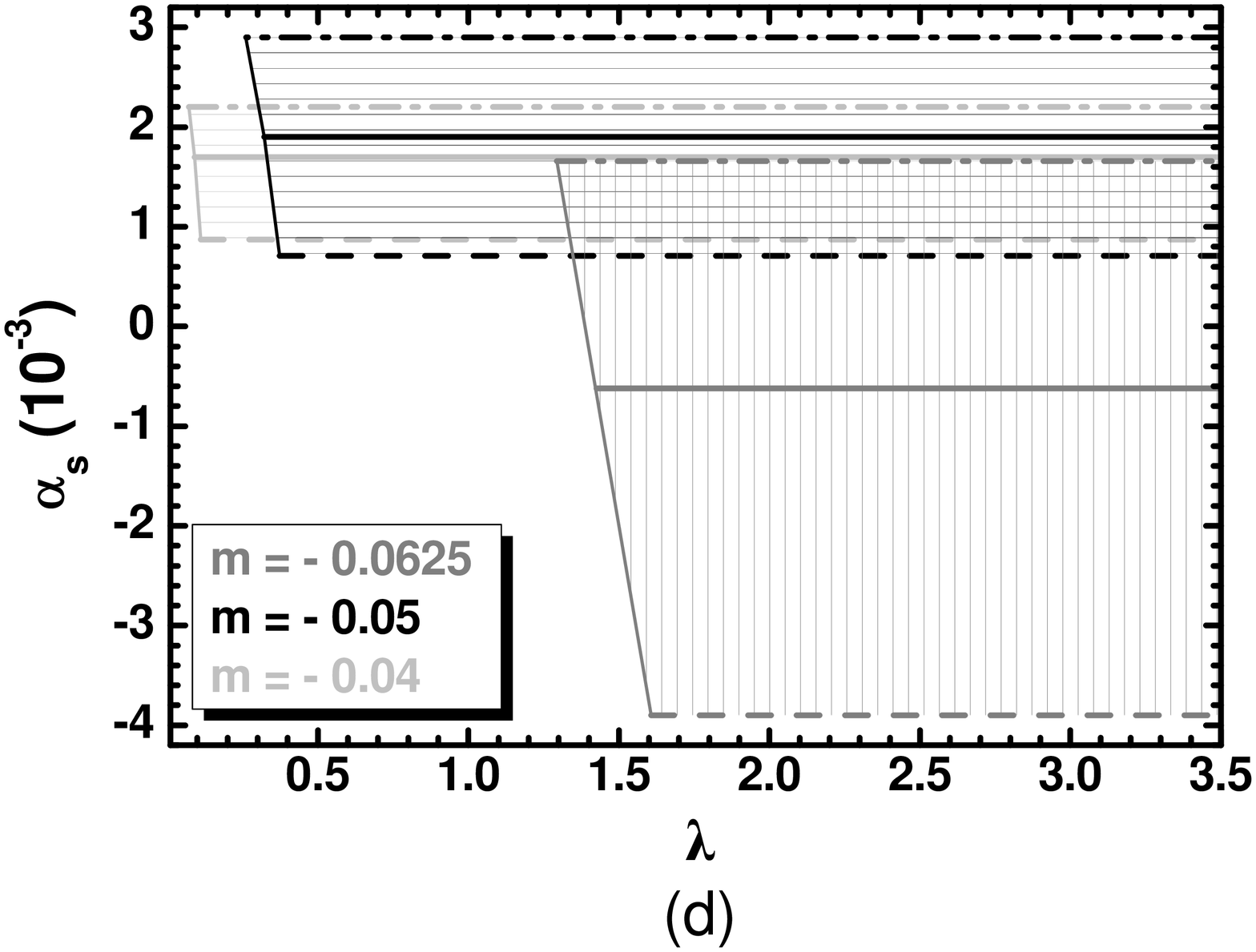,height=3.6in,angle=-90} \hfill
\end{minipage}
\hfill \vchcaption[]{\sl\small The (hatched) regions allowed by
Eqs.~(\ref{Nhi}), (\ref{Prob}), (\ref{nswmap}{\sf\ftn a, b}) and
(\ref{subP}) in the $\ld-\ck$ ({\sf\ftn a}), $\ld-\ksp$ ({\sf\ftn
b}), $\ld-r$ ({\sf\ftn c}), $\ld-\as$ ({\sf\ftn d}) plane  for
$n=2$, $\ks=0.1$, $\kpp=0.5$ and $m=-0.04$ (light gray lines and
hatched regions), $m=-0.05$ (black lines and hatched regions)
$m=-0.0625$ (gray lines and hatched regions). The conventions
adopted for the type and color of the various lines are also shown
in the label of panel {\sf\ftn (a)}.}\label{fig2gm}
\end{figure}


Confronting the models under consideration with the constraints of
Eqs.~(\ref{Nhi}), (\ref{Prob}), (\ref{nswmap}{\sf\ftn a, b}) and
(\ref{subP}) we depict the allowed (hatched) regions  in the
$\ld-\ck$, $\ld-\ksp$, $\ld-r$ and $\ld-\as$ plane for $m=-0.04$
(light gray lines and horizontally hatched regions), $m=-0.05$
(black lines and horizontally hatched regions), $m=-0.0625$ (gray
lines and vertically hatched regions) -- see \sFref{fig2gm}{a},
{\sf\ftn (b), (c)} and {\sf\ftn (d)} respectively. In the
horizontally hatched regions $r$ is compatible with \Eref{rgw}
whereas in the vertically hatched region $r$ turns out to be close
to the central value suggested \cite{gws} by \bicep\ -- after
subtraction of a dust foreground. The conventions adopted for the
various lines are also shown in the label of panel {\sf\ftn (a)}.
In particular, the dashed [dot-dashed] lines correspond to $n_{\rm
s}=0.975$ [$n_{\rm s}=0.946$], whereas the solid (thick) lines are
obtained by fixing $n_{\rm s}=0.96$ -- see \sEref{nswmap}{a}. The
lower bound for the regions presented in \Fref{fig2gm} is provided
by the constraint of \sEref{subP}{b} which is saturated along the
thin lines. The perturbative bound on $\ld$ limits the various
regions at the other end.

From \sFref{fig2gm}{a} we remark that $\ck$ remains almost
proportional to $\ld$ but the dependence on $\ksp$ is stronger
than that shown in \rFref{2}--{\sf\ftn (a$_1$)} and {\sf\ftn
(a$_2$)}. Also as $|m|$ increases, the allowed areas become
smaller favoring larger $\ck$'s and $\ld$'s. From
\sFref{fig2gm}{b} we notice that the allowed $\ksp$'s get
concentrated around zero as $|m|$ increases and so the relevant
tuning increases. Finally from \sFref{fig2gm}{c} and {\sf\ftn (d)}
we conclude that decreasing $m$ below zero, $r$ and $\as$ increase
w.r.t their standard values -- cf. \rEref{3.22} and discussion
below \srEref{3.32}{c}. As a consequence, $r$ for $m=-0.04$ and
$-0.05$ approaches the range of \Eref{rgw} -- which explains
(conservatively) the recent \bicep\ results -- being at the same
time compatible with the \plk\ (and WMAP) measurements. For
$m=-0.0625$, $r$ reaches its (almost) maximal possible value in
our set-up which lie close to the \bicep\ central $r$ value -- see
\Sref{intro} above \Eref{rgw}. On the other hand, $\as$ remains
sufficiently low; it is thus consistent with the fitting of data
with the standard $\Lambda$CDM model -- see \srEref{3.6}{b}.
Namely, $|\as|$ never exceeds $4\cdot10^{-3}$ and it is mostly
positive. It is clear, therefore, that it is much smaller than its
best-fit value of roughly $-0.02$ which may help \cite{gws,boyle}
to relieve the tension between the \bicep\ and the \plk\ data as
regards the bounds on $r$. Furthermore, the resulting  $\as$
follows the behavior of $r$, which depends only on the input $m$
and $\ksp$ (or $\ns$) and are independent on $\ld$ (or $\ck$) --
as anticipated in the end of \Sref{gan}. More explicitly, for
$n_{\rm s}=0.96$ and $\Ne_\star\simeq55-57$ we find:
\beqs\bea\label{resgm} && 0.17\lesssim {\ck\over
10^4}\lesssim6.7\>\>\>\mbox{with}\>\>\>0.09\lesssim
\ld\lesssim3.5\>\>\>\mbox{and}\>\>\> 0.2\lesssim
-{\ksp\over0.1}\lesssim9.3\, \>\>\>(m=-0.04);\>\>\>\>\>\>\\ &&
\label{resgm2} 0.56\lesssim {\ck\over
10^4}\lesssim6.1\>\>\>\mbox{with}\>\>\>0.32\lesssim
\ld\lesssim3.5\>\>\>\mbox{and}\>\>\> 0.02\lesssim
-{\ksp\over0.1}\lesssim2.2\,\>\>\>(m=-0.05);\>\>\>\>\>\>\>\>\>\>\>\>\>\\
&& \label{resgm3}2.6\lesssim {\ck\over
10^4}\lesssim6.45\>\>\>\mbox{with}\>\>\>1.4\lesssim
\ld\lesssim3.5\>\>\>\mbox{and}\>\>\> 1.9\lesssim
-{\ksp\over0.01}\lesssim4.7\,\>\>\>(m=-0.0625).\>\>\>\>\>\>\>\>\>\>\>\>\>
\eea\eeqs
In these regions we obtain
\beq \label{resgm4}  {r\over0.1}=0.53, 0.96,
1.6\>\>\>\mbox{and}\>\>\>
{\as\over0.001}=1.7,1.9,-0.6\>\>\>\mbox{for}\>\>\>
-{m\over0.01}=4,5,6.25\eeq respectively. Consequently, our model
can fit both \plk\ and \bicep\ results adjusting just two more
parameters ($m$ and $\ksp$) than those employed in the initial
(and more robust) model \cite{pallis} exhibiting the no-scale-type
symmetry.

It is worth noticing that a decrease of $\ksp$ below zero is
imperative in order to achieve a simultaneous fulfillment of
\sEref{nswmap}{a} and (\ref{rgw}). Indeed, selecting $\ksp=0$ the
increase of the prefactor $(-3)$ in $K$ generates an enhancement
of $r$ which is accompanied by an increase of $\ns$ beyond the
range of \sEref{nswmap}{a}. Therefore, the new solutions to the
tachyonic instability problem which avoid terms of the form
$|S|^2|\Phi|^2$ in $K$ \cite{ketov, noK, nil} are expected not to
fit well with our proposal. Increasing, finally, $n$ above $2$ the
required $\ld$ and $\ck$ values become larger and so the allowed
regions are considerably shrunk; we thus do not pursue further our
investigation.

In closing, it would be instructive to compare our proposal with
the so-called $\alpha$-attractor models \cite{aroest} where
deviations from the conventional ($-3$) coefficient of the
logarithm in the \Ka\ are also investigated. Namely, focusing in
Sec.~7.2 of the second paper in Ref.~[12] we can remark the
following essential differences:

\begin{itemize}

\item[{\sf\ftn (i)}] In the \Ka\ the inflaton appears linearly,
and not quadratically as in our case, without a large coefficient
$\ck$ and no terms exist proportional to $\ksp$ and $\kpp$.
Therefore, no dependence on those parameters is studied. Moreover
no restrictions from IG are taken into account.

\item[{\sf\ftn (ii)}] The numerical prefactor of the logarithm in
the \Ka\ appears also in the exponent of the superpotential in a
such way that the inflationary potential, derived from
\Eref{Vhig}, has no dependence on $\alpha=1+m$ besides the one
involved in expressing the JF inflaton $T$ in terms of EF one
$\varphi$. The inflationary potential depends only on an
\emph{arbitrary} exponent called $n$ which enters the definition
of the function $\tilde f$ in Eq.~(7.12). In an explicit example
mentioned in the last paragraph of Sec.~7.2 the form $\tilde
f=T-1$ is adopted and the inflationary potential has the simplest
form $V_0(1-e^{-\sqrt{2/3\alpha}\varphi/\mP})^2$.

\end{itemize}

As a consequence of the arrangements above, the models of
\cref{aroest} asymptote to Starobinsky model for low $\alpha$'s
and to quadratic inflation for very large $\alpha$'s -- obviously,
trasplanckian values for $T$ and $\varphi$ are employed. This
behavior is not observed in our setting. The reason can be
transparently shown if we express $\Vhio$ in \Eref{3Vhiom} in
terms of $\se$ defined in \Eref{se1} -- with $\se_{\rm c}=0$ -- as
follows:
\beq
\label{Vse}\Vhio\simeq\frac{\ld^2\mP^4}{4\fsp\ck^2}e^{-\sqrt{6/(1+m)}m\se/\mP}\lf1-e^{-\sqrt{2/3(1+m)}\se/\mP}\rg^2,\eeq
where we assumed $f_{\phi\phi}\simeq1$ and the last factor
originates from the quantity $f_2^2/\ck^2\xsg^4$ -- with $\fpp$
defined below \Eref{3Vhioo}. From the last expression we can
easily infer that, for $m\neq0$, $\Vhio$ declines away from the
simplest form found in \cref{aroest}. Indeed,  in our set-up the
last factor of \Eref{Vse}, which already exists in \cref{aroest},
is multiplied by
$\xsg^{-6m}/\fsp\ck^{3m}=e^{-\sqrt{6/(1+m)}m\se/\mP}/\fsp$. This
last factor has a significant impact on our results -- see
\Eref{ragm}.

\section{The Effective Cut-off Scale}\label{fhi3}

The realization of IG inflation with $m<0$ retains the
perturbative unitarity up to $\mP$ as the models described in
\cref{pallis} do -- cf.~\cref{riotto, R2r}. Focusing first on the
JF computation, we remark that the argument goes as analyzed in
\rSref{4.1} with $F_{\rm K}$ taking the form
\beq F_{\rm
K}\simeq1-{nm\over(1+m)}+\frac{3}{2\xsg^2}\:mn^2\fk\>\Rightarrow\>\vev{F_{\rm
K}}\simeq1-{nm\over(1+m)}+\frac382^{2/n}\:mn^2\ck^{2/n},\label{Fk}\eeq
as can be easily inferred from the second term of \Eref{Sfinal}.
The last expression in \Eref{Fk} can be extracted taking into
account  \eqs{vevs}{se1}. Here, and henceforth, we keep the
dependence of the formulas on the exponent $n$ for better
comparison with the formulas in \rSref{4}. Inserting \Eref{Fk}
into \rEref{4.3} we can conclude that UV cut-off scale $\Qef$ is
still roughly equal to $\mP$ since the dangerous prefactor
$\ck^{-2/n}$ is eliminated. Needless to say, terms proportional to
$\ksp$ or $\kpp$ included in \Eref{Kolg} are small enough and do
not generate any problem with the perturbative unitarity.
Therefore, they do not influence our conclusions.

Moving on to the EF, recall -- see \eqs{Jg}{se1} -- that the
canonically normalized inflaton,
\beq\dphi=\vev{J}\dph\>\>\>\mbox{with}\>\>\>\vev{J}\simeq\sqrt{\frac{3(1+m)}{2}}\frac{n}{\vev{\xsg}}=
{\sqrt{3}\over2}\ n\sqrt[n]{2\ck} \label{dphi} \eeq
acquires mass which is calculated to be
\beq \label{masses} \msn=\left\langle\Ve_{\rm
IG0,\se\se}\right\rangle^{1/2}\simeq\ld\mP/\sqrt{3(1+m)}\ck\,.\eeq
We remark that $\msn$ turns out to be largely independent of $n$
as in \rEref{4.5}. However, due to the modified  $\ld-\ck$
relation -- see \Eref{langm} -- and the factor $\sqrt{1+m}<1$ in
the denominator, its numerical value increases slightly w.r.t its
value in the models of \cref{pallis}. E.g., taking
$\sg_\star=0.6\mP$ and $m=-(0.04-0.625)$ we get
$6.9\lesssim\msn/10^{13}~\GeV\lesssim9.2$ for $\ns$ in the range
of \srEref{3.6}{a}. Since we do not find any attractor towards the
quadratic inflation, $\msn$ is clearly disguisable from its value
encountered in that model.

To check the limit of the validity of the effective theory, we
expand $J^2 \dot\phi^2$ involved in \Eref{Saction1} about
$\vev{\phi}$ in terms of $\dphi$ in \Eref{dphi} and we arrive at
the following result
\beq\label{exp2} J^2
\dot\phi^2=\lf1-\frac{2(1+2m+m^2)}{n(1+m)^{5/2}}\sqrt{\frac{2}{3}}\frac{\dphi}{\mP}+\frac{2}{n^2(1+m)}\frac{\dphi^2}{\mP^2}-
\frac{8}{3n^3(1+m)^{3/2}}\sqrt{\frac{2}{3}}\frac{\dphi^3}{\mP^3}+\cdots\rg\dot\dphi^2.\eeq
The expansion corresponding to $\Vhio$ in \Eref{3Vhioo}  with
$\ksp\simeq0$ and $\kpp\simeq0$ includes the terms -- cf.
\srEref{4.6}{c}:
\bea\nonumber
\Vhio=\frac{\ld^2\mP^2\dphi^2}{6\ck^2(1+m)}&\cdot&\Bigg[1-\sqrt{\frac{2}{3}}\lf1+{1\over
n}+m\lf4+3m+{1\over n}\rg\rg\frac{\dphi}{(1+m)^{3/2}\mP}\\
&&+\lf\frac{7}{18}+{1\over n}+\frac{11}{18n^2}+m\lf2+3m+{3\over
n}\rg\rg\frac{\dphi^2}{(1+m)\mP^2}-\cdots\Bigg]\cdot\>\>\>\label{Vexp}\eea
Hence, we can conclude from Eqs.~(\ref{exp2}) and (\ref{Vexp})
that in this case also $\Qef=\mP$, in agreement with our analysis
in the JF.

\section{Conclusions}\label{con}

Prompted by the recent excitement -- see e.g.
\cref{rEllis,rStar,aroest} -- in the are(n)a of inflationary model
building, we carried out a confrontation of IG inflation,
formulated beyond the no-scale SUGRA, with the \plk\ \cite{plin}
and \bicep\ results \cite{gws} -- regardless of the ongoing debate
on the ultimate validity of the latter \cite{gws1,gws2}. As in our
original paper, \cref{pallis}, the inflationary models are tied to
a superpotential, which realizes easily the idea of IG, and a
logarithmic \Ka, which includes all the allowed terms up to the
fourth order in powers of the various fields -- see \Eref{Kolg}.
The models are totally defined imposing two global symmetries -- a
continuous $R$ and a discrete $\mathbb{Z}_n$ symmetry -- in
conjunction with the requirement that the original inflaton takes
\sub\ values. Extending our work in \cref{pallis} we allow for
deviations from the prefactor $(-3)$ multiplying the logarithm of
the \Ka -- see \Eref{Kolg}. We parameterized these deviation by a
factor $(1+m)$. Fixing $n=2$, confining $m$ to the range
$-(4-6.25)\%$ and adjusting $\ld,~\ck$ and $(-\ksp)$ in the ranges
$0.09-3.5$, $(1.7-64.5)\cdot10^3$ and $0.019-0.93$
correspondingly, we achieved inflationary solutions that are
simultaneously \plk\- and \bicep -friendly, i.e. we obtained
$\ns\simeq0.96$ and $0.05\lesssim r\lesssim0.16$ with negligible
small $\as$. A mild tuning of $\kx$ to values of order $0.1$ is
adequate such that the one-loop radiative corrections remain
subdominant. Moreover, the corresponding effective theory remains
trustable up to $\mP$, as in the other cases analyzed in
\cref{pallis}. As a bottom line we could say that although
incarnations of IG inflation beyond the no-scale SUGRA are less
predictive than the simplest model presented in \cref{pallis} they
provide us with the adequate flexibility needed to obtain larger
$r$'s without disturbing the remaining attractive features of this
inflationary model.

\begin{acknowledgement}

$~$ This research was supported from the Generalitat Valenciana
under contract PROMETEOII/2013/017. I~would like to acknowledge
useful discussions with G.~Lazarides and dedicate this paper to
the memory of my High School Greek language teacher V.~Aspiotis, a
friend and a mentor with a valuable impact on my culture.

\end{acknowledgement}

\end{document}